# Continuously Tunable Acoustic Metasurface with Rotatable Anisotropic Three-component Resonators

Pan Li, Yunfan Chang, Qiujiao Du, Zhihong Xu, Meiyu Liu and Pai Peng

School of Mathematics and Physics, China University of Geosciences, Wuhan 430074, China

*Corresponding author. paipeng@cug.edu.cn

**Abstract**

We propose a tunable acoustic metasurface consisting of identical units. And units are rotatable anisotropic three-component resonators, which can induce the non-degenerate dipolar resonance, causing an evident phase change in low frequencies. Compared with the monopole resonance widely used in Helmholtz resonators, the polarization direction of the dipole resonance is a new degree of freedom for phase manipulation. The proposed metasurface is constructed by identical units that made with real (not rigid) materials. And the phase profile can continuously change by rotating the anisotropic resonators. We present a wide-angle and broad-band acoustic focusing by the metasurface under a water background.

## 1. Introduction

Acoustic metasurfaces (AMs) can freely manipulate wavefront have received extensive attention in recent years[1-4]. Most of the classical acoustic metasurfaces (CAMs) have two ways to modulate the phase change inside the unit cells. One way is to change the effective spatial path of waves by the space-coiling structures[2, 5, 6], and the other is to obtain phase delay from resonances[7, 8]. These CAMs have realized many interesting functions, including abnormal reflection[5] or refraction[9], acoustic focusing[6], asymmetric transmission[10, 11], acoustic accelerating beam[8, 12], acoustic holography[13], acoustic illusion[14] and cloaking[15]. Although CAMs hold great potentials for wavefront manipulation, they are limited to meet the requirements of broadband and alterable functionalities due to fixed microstructures. To overcome this limitation, researchers propose the tunable acoustic metasurfaces (TAMs)[16-27], whose acoustic response can be modulated by tunable units, like actively controlled transducer arrays, independently adjustable units or reconfigurable microstructures.

Currently, there are two main types of TAMs: active and passive. And most of the TAMs still rely on actively controlled units, like transducer arrays[16, 17] and active membrane metamaterials[19, 20], which are complex and expensive. Recently, researchers have paid more attention to passive TAMs consisting of reconfigurable microstructures[22-27]. Based on the matched screw-and-nut physical mechanism, Zhao et al. designed a class of the tunable space-coiling metasurface with individual unit components of a helix screwed inside a plate for transmitted[24] and reflected wavefront modulation[25]. With the nested Helmholtz resonant structure, Zhai et al.

designed a TAM for filtering and imaging[22]. And Wang et al. proposed a TAM consisting of annular resonators to modulate the transmitted wavefront[27]. Besides, a TAM composed of tunable Helmholtz resonators has been reported by Tian et al.[23]. So far, the research works about TAM are still limited.

In this paper, we design a tunable acoustic metasurface (TAM) based on identical anisotropic resonant units[28-30], each of which is a modified three-component composite. The anisotropic resonant unit is proven to be a dipolar resonator with two non-degenerate eigenmodes, which could be controlled by the rotation angle of the elliptical rotor. And the reflected phase shift of the unit will have a span of $2\pi$ just by changing its rotation angle. In a wide frequency range, this TAM can focus the reflected waves of different incident angles on a fixed focal length. With the same TAM, we also can control the position of the focus point arbitrarily.

## 2. Design of the Tunable Unit Cell

We consider a tunable unit cell, which consists of a square epoxy frame with a circle cavity and an elliptical resonator in the center of the cavity, as shown in Fig. 1(a). The length of the frame and the radius of the cavity are $p$ and $r_3 = 0.4p$, respectively. The cavity is filled up with water, so the resonator is rotatable. The resonator is particular designed with three components[31] to provide the dipolar resonance[28]. The resonator is composed of an elliptical epoxy shell (with semi-minor axis $R_1 = 0.3p$ and semi-major axis $R_2 = 0.35p$), an elliptical rubber layer (with semi-minor axis $r_1 = 0.25p$ and semi-major axis $r_2 = 0.3p$), and a circle steel core

(with radius $r_0 = 0.18p$). The rotational angle of resonator is $\theta$, which is the only variable parameter in our system and can continuously change from 0 to 90 degree. The used material parameters are: $\rho_e = 1180 \text{kg/m}^3$, $\lambda_e = 4.4 \times 10^9 \text{N/m}^2$, and $\mu_e = 1.6 \times 10^9 \text{N/m}^2$ for epoxy; $\rho_w = 1000 \text{kg/m}^3$ and $c_w = 1490$ m/s for water, $\rho_r = 980 \text{kg/m}^3$, $\lambda_r = 1.96 \times 10^9 \text{N/m}^2$, and $\mu_r = 5.5 \times 10^5 \text{N/m}^2$ for rubber; $\rho_s = 7900 \text{kg/m}^3$, $\lambda_s = 1 \times 10^{11} \text{N/m}^2$, and $\mu_s = 8.1 \times 10^{10} \text{N/m}^2$ for steel. Here $\rho$ is the mass density, $\lambda$ and $\mu$ are the Lamé constants, and $c$ is the speed of sound. An elastic metamaterial is constructed by the tunable unit cells (with $\theta = 0°$) periodically arranged in a square lattice. We calculate the band structure with the finite element method (using COMSOL Multiphysics software) and plot the lowest seven bands in Fig. 1(b). By carefully check the patterns of eigenstates, we find three flat bands induced by rotational resonances (denoted by blue hollow circles around $p/\lambda = 0.026$, $p/\lambda = 0.047$ and $p/\lambda = 0.059$), and the other four bands (highlighted by red solid lines) are induced by a non-degenerate dipolar resonance[28]. The eigenstates of the eigenmodes A and B on the fifth and third bands (with normalized wavelengths $p/\lambda_A = 0.071$ and $p/\lambda_B = 0.052$) on the Brillouin zone boundary along the $\Gamma Y$ direction are plotted in Fig. 1(c) and (d), respectively. The movements of modes A and B are mainly along the semi-major and semi-minor axis of the ellipse, respectively. These two modes with movement perpendicular to each other are a pair eigenmodes of a non-degenerate dipolar resonance. The corresponding wavelengths of mode A and B are $\lambda_A = 14.1p$ and $\lambda_B = 19.2p$, respectively. The modes A and B present the longitudinal and transversal modes of a dipolar resonance, respectively, due to their wavelengths.

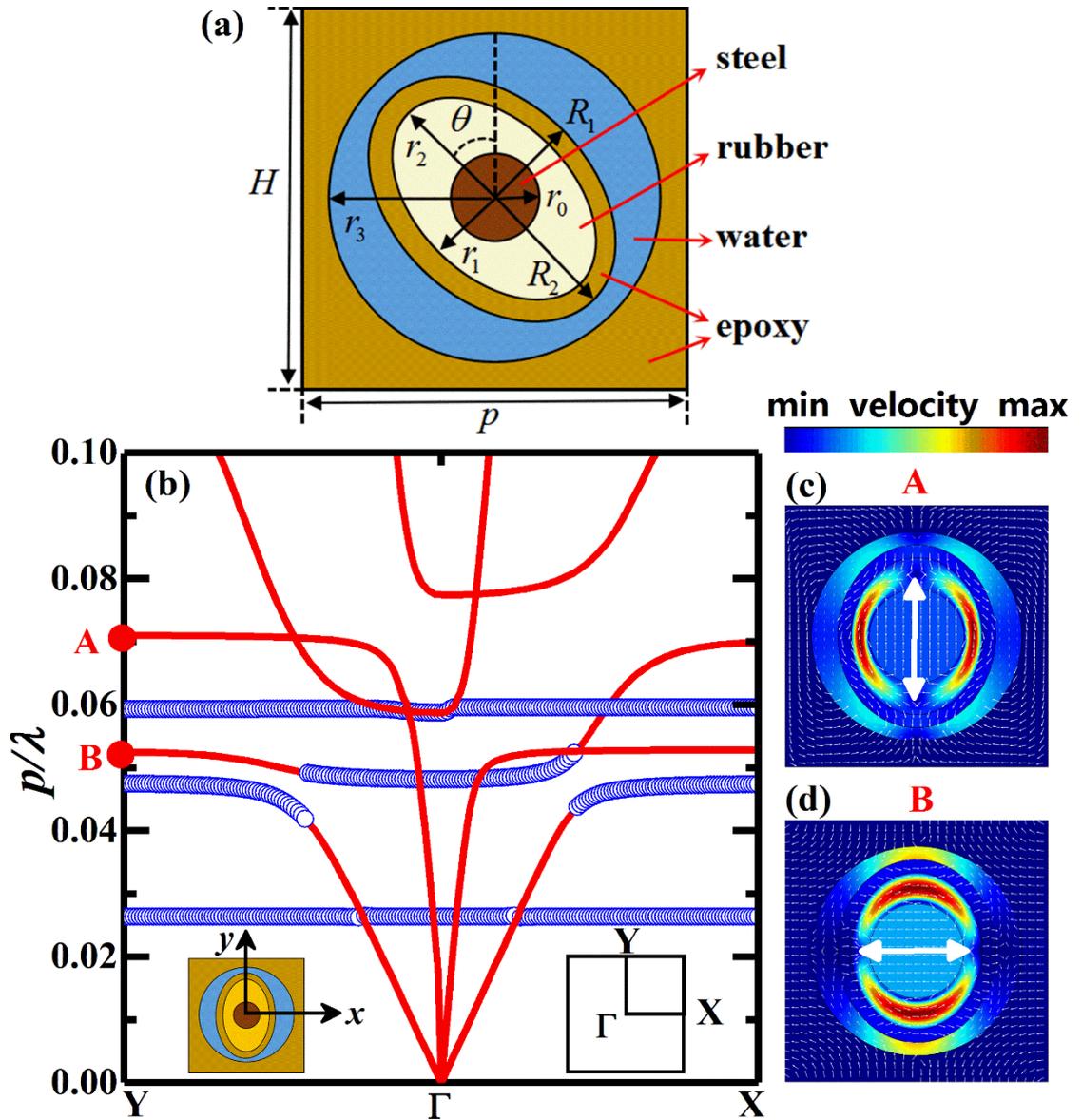

**FIG. 1** (a) The schematic diagram of the tunable unit cell. (b) The band structure of the elastic metamaterial, which is composed of the tunable unit cells (with $\theta = 0°$) periodically arranged in a square lattice. (c) and (d) respectively are the excited velocity fields of points A (with $p/\lambda_A = 0.071$) and B (with $p/\lambda_B = 0.052$) denoted in (b).

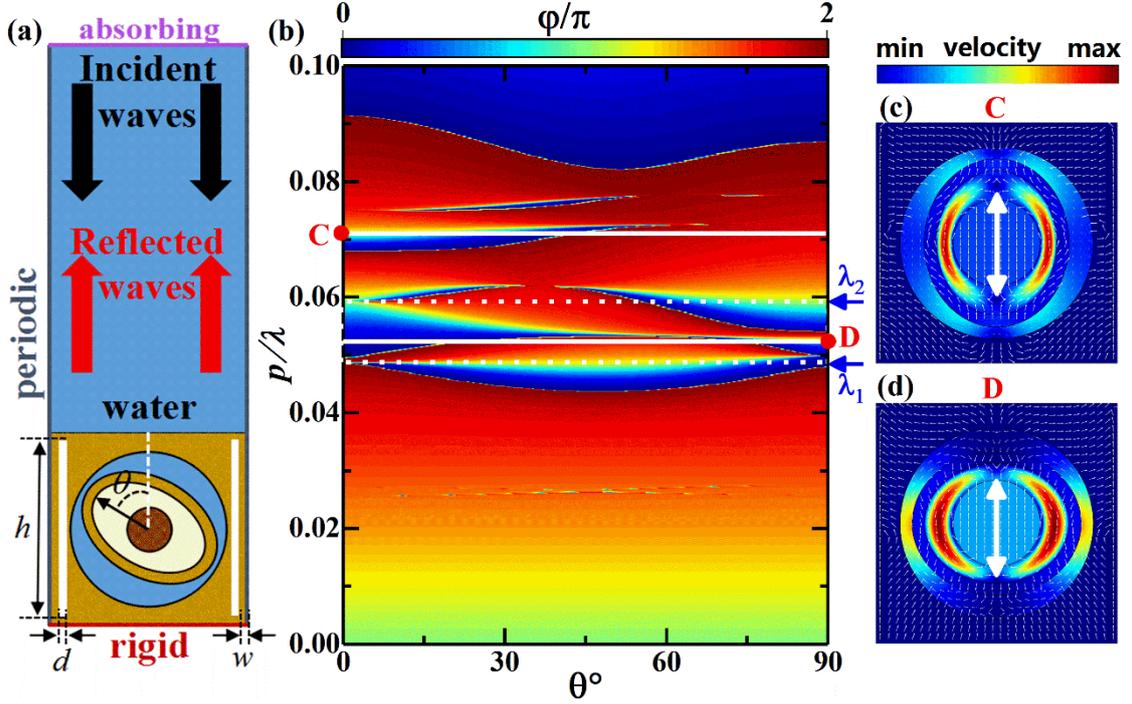

**FIG. 2** (a) The calculation region of the TAM. Plane pressure waves are normally incident from water. (b) The reflected phase change of the tunable unit cell as a function of rotational angle and wavelength. (c) and (d) are the excited velocity fields of points C (with $p/\lambda_A = 0.071$) and D (with $p/\lambda_B = 0.052$) denoted in (b), respectively.

We let one layer of the elastic metamaterial lied at the bottom of water and study the phase properties for reflected waves. The calculation region is shown in Fig. 2(a). In particular, to reduce the interaction between the neighboring units, two air voids (with length $d = 0.02p$ and the thickness $h = 0.96p$) are added on the left and right sides of the unit. The distance between the air void and the side boundary is $w = 0.01p$. The used material parameters are $\rho_a = 1.29 \text{kg/m}^3$ and $c_a = 340 \text{m/s}$ for air. We let plane pressure waves normally incident from water and calculate the reflected phase. The results are plotted in Fig. 2(b), where the phase change $\varphi$ is as a function of the rotational angle $\theta$ and incident wave length $\lambda$. In general, the phase change $\varphi$ is trivially very small and insensitive to the rotated angle because of long wavelengths, which are more than 10 times the unit size. However, large phase changes are found in a range between $p/\lambda = 0.045$ and $p/\lambda = 0.08$. We choose a point C, which has an incident wavelength of $\lambda_A$ the same as that of mode A, and plot the excited velocity field in Fig. 2(c). The vibration is mainly along the semi-major axis, and the pattern of point C is almost the same as that of mode A. We can see that the longitudinal mode of the dipolar resonance is excited by the incident acoustic wave, and the large phase change around the point C is induced by the longitudinal mode. Here the elliptical resonator is vertically placed and the rotational angle is $\theta = 0°$. If the elliptical resonator is horizontally placed (corresponding to $\theta = 90°$), the longitudinal mode cannot be excited due to symmetry, and thus the phase change will reduce to trivially small. As a result, the phase change $\varphi$ is controlled by the rotational angle $\theta$. When $\theta$ gradually increases from $\theta = 0°$ to $\theta = 90°$, the phase change $\varphi$ can roughly cover a range of $2\pi$. Base on this phase properties, we can design an AM consisting of identical until cells, and the rotational angle $\theta$ is a new degree of freedom to control the phase change. A continuously tunable rotational angle brings a flexible phase regulation, resulting in a continuously TAM.

Similar phase properties can be found at the incident wavelength of $\lambda_B$. When the elliptical resonator is horizontally placed (with $\theta = 90°$), the transverse mode of the dipolar resonant can be excited. The point D marked in Fig. 2(b) has an incident wavelength of $\lambda_B$ the same as that of mode B. The excited velocity field of point D is plotted in Fig. 2(d). The vibrations are mainly along the semi-minor axis, where the pattern is the same as that of mode B. In contrast to the longitudinal mode, the transverse mode can induce large phase change at $\theta = 90°$ but trivially small phase change at $\theta = 0°$. The two (longitudinal and transverse) resonant modes of a non-degenerate dipolar resonance have different resonant wavelengths, which make the dependent relationship between phase change and rotational angle exist at a broad wavelength range. In a range roughly between $p/\lambda = 0.045$ and $p/\lambda = 0.08$, the phase changes are controlled by the rotational angle. Not all the wavelength, the phase change $\theta$ can cover a full $2\pi$ span. The refined wavelengths are in a region between $p/\lambda_1 = 0.049$ and $p/\lambda_2 = 0.059$ (corresponding to wavelengths $\lambda_1 = 20.4p$ and $\lambda_2 = 16.9p$) as denoted in Fig. 2(b) by two white dashed lines. In this region, the phase change $\theta$ can cover a full $2\pi$ span, and thus we can build a continuously broad-band TAM.

## 3. The Continuously Tunable Acoustic Metasurface

As shown in Fig. 3(a), a TAM is composed of 120 identical tunable unit cells as studied in Fig. 2(a). We let pressure waves incline incident (with incident angle $\alpha$) from the water and manipulate the reflected wavefront by using the TAM. In this work, we demonstrate that the TAM can focus (as an example) the reflected waves and manipulate the focus position by adjusting the rotational angle in every unit cell when the incident wave come from different directions and with different wavelengths.

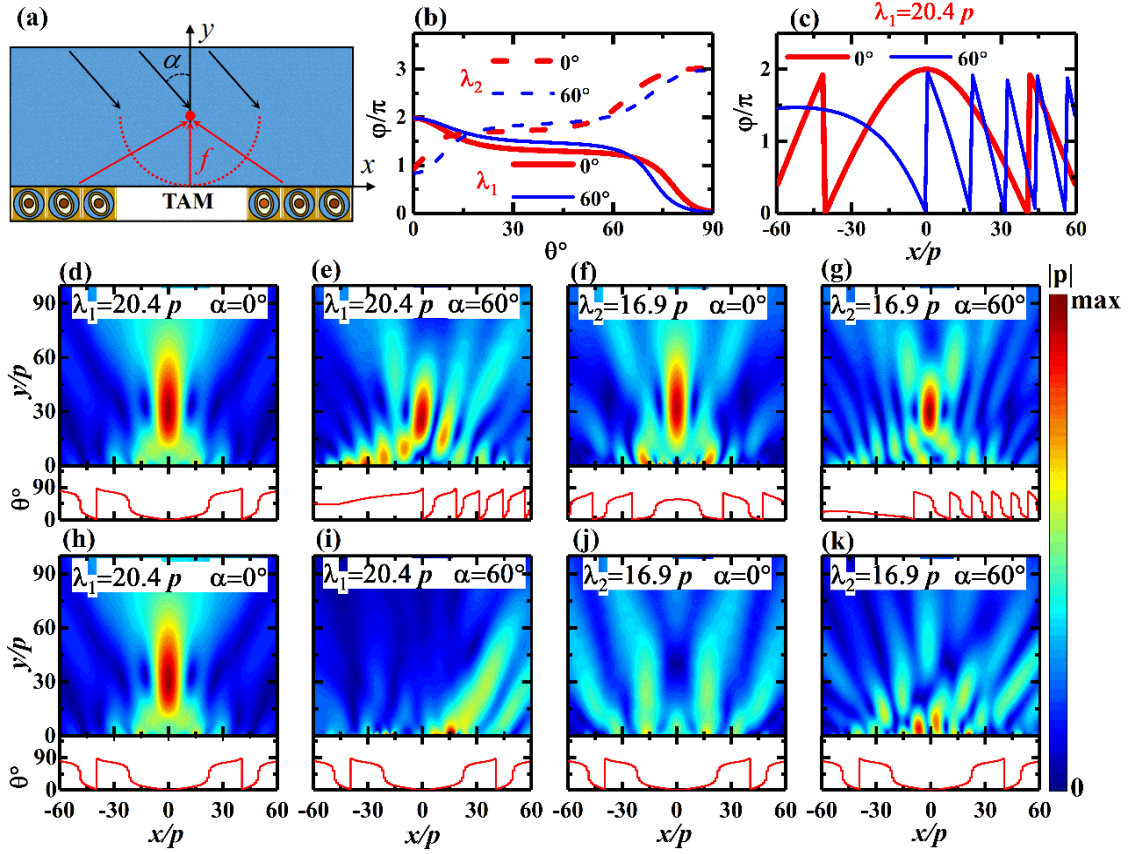

**FIG. 3** The TAM with a fixed focal length of $f = 30p$. (a) The schematic diagram of the TAM composed of 120 tunable unit cells. (b) The phase change is as a function of the rotational angles $\theta$. (c) The required phase profile obtained from Eq. (1) (d-g) The reflected amplitude fields (up) and the rotational angles $\theta$ of the unit cells (down) for different incident angles and wavelengths. (h-k) The reflected amplitude fields obtained by a TAM with fix geometry.

Base on the Generalized Snell's Law (GSL)[32], the reflected pressure field is determined by the phase profile $\varphi(x)$ provided by the TAM. A focusing effect

requires a hyperbolic function of $\varphi(x)$. The expression is

$$\varphi(x) = \frac{2\pi}{\lambda}(\sqrt{(x-x_0)^2 + f^2} - f - x\sin\alpha). \tag{1}$$

Here we fix the focal length of $f = 30p$ at $x_0 = 0$, and the phase profile $\varphi(x)$ can be calculated with given parameters of incident angle $\alpha$ and wavelength $\lambda$. Following the relationship between phase change and rotational angle $\varphi(\theta)$, which has been calculated in Fig. 2(b), the rotational angle distributions of the unit cells $\theta(x)$ can be achieved to satisfy the required phase profile $\varphi(x)$. For example, in case of $\alpha = 0°$ and $\lambda = \lambda_1$, the functions of $\varphi(\theta)$ and $\varphi(x)$ can be obtained from Fig. 2(b) (the lower white dashed line) and Eq. (1), and the results are plotted in Fig. 3(b) and (c) by the red thick lines, respectively. Then the rotational angle distributions $\theta(x)$ is obtained from $\varphi(\theta)$ and $\varphi(x)$, and the result is plotted at the bottom of Fig. 3(d). The corresponding reflected pressure field is shown in Fig. 3(d). The pressure pattern exhibits a focal point at about $y = 30.6p$, which agrees well with the prediction ($f = 30p$). When the incident angle changes to $\alpha = 60°$, the acoustic focusing should fail, as shown in Fig. 3(f) as a reference, if the geometry of TAM is fixed as CAMs. Here we show that the TAM can adapt to the change of incident direction by adjusting the rotational angle distributions. In case of $\alpha = 60°$ and $\lambda = \lambda_1$, the functions of $\varphi(\theta)$ and $\varphi(x)$ are recalculated and plotted in Fig. 3(b) and (c) by the blue thin lines, respectively. Accordingly, the obtained $\theta(x)$ as well as the pressure field are shown in Figs. 3(e). The focal point is kept at $y = 27.4p$, which is slightly different to the designed position. Good focusing effects can be achieved by the TAM when the incident angle changes within a range $-60° \leq \alpha \leq 60°$. When the incident wavelength changes to $\lambda = \lambda_2$, the TAM can also keep the focus position within a range $-60° \leq \alpha \leq 60°$. Figures 3(f) and (g) show results corresponding to the cases of normal ($\alpha = 0°$) and incline ($\alpha = 60°$) incidences at the wavelength of $\lambda = \lambda_2$, respectively. The rotational angle distribution $\theta(x)$ is obtained in a similar way, and the pressure field also shows good focusing effects. Figures 3(j) and (k) are the results obtained from the AM with fixed geometry.

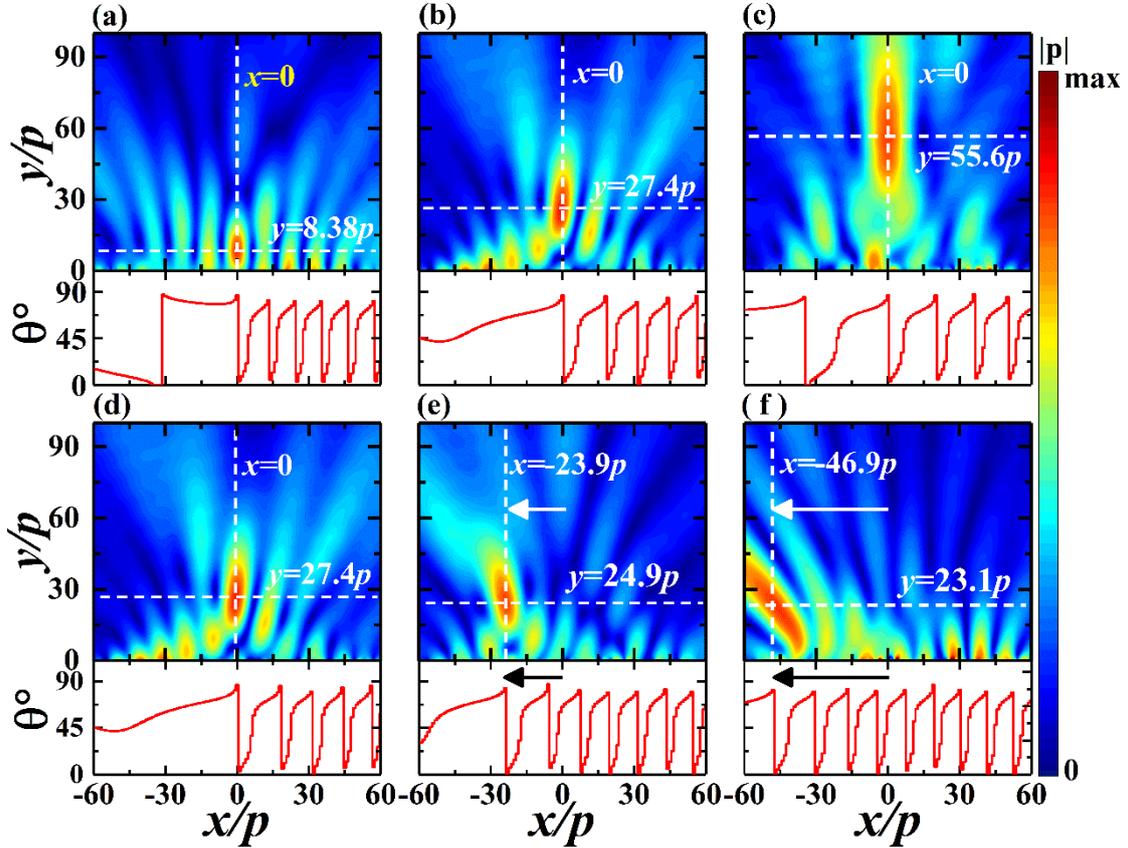

**FIG. 4** The TAM with an adjustable focus position. The reflected amplitude fields (up) and the rotational angles $\theta$ of the unit cells (down) at the wavelength of $\lambda_1 = 20.4p$ and the incident angle of $\alpha = 60^0$. The black arrow denotes the phase shift.

Next, we show the manipulation of focus position. Without loss of generality, we set the incident angle $\alpha = 60°$ and wavelength of $\lambda = \lambda_1$ as constants. The manipulation of vertical and horizontal position can be achieved by changed the focal length and shift the rotational angle distributions, respectively. Different focal length $f$ lead to different phase profile $\varphi(x)$. The rotational angle distributions $\theta(x)$ can be obtain form the $\varphi(\theta)$ studied in Fig. 3(b) and a new function of $\varphi(x)$ from Eq. (1). Figures 4(a), (b) and (c) show the pressure fields with focal points at $y = 8.38p$, $y = 24.9p$ and $y = 55.6p$, which are roughly agree with the designed focal lengths of $f = 10p$, $f = 30p$ and $f = 60p$, respectively. Figures 4(d)~(f) show that the focus position shift to the left side when we shift the rotational angle distribution. The focusing effect becomes weaker because of the edge effect. In principle, an infinite long TAM can smoothly move the whole pressure field along the horizontal direction. If we simultaneously change the focal length and shift the rotational angle distribution, we can manipulate the focus position in a wide range.

## 4. Conclusion

In summary, we propose a TAM composed of identical anisotropic resonant units, which can induce the non-degenerate dipolar resonance, causing an evident phase change. With anisotropic resonant units, we can control phase changes with a new degree of freedom, the polarization of the dipole resonance. As results show, the TAM can focus acoustic waves at a fixed length when they incident from different directions with different wavelengths, and can manipulate the focus position. Besides, the TAM is made of actual materials and without any rigid material. Here, the acoustic focusing effect is presented under a water background. The proposed wide-angle and broad-band TAM may have good potential in the application.

**Acknowledgement**

This work was supported by the National Natural Science Foundation of China (Grant No: 11604307).